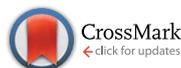

# Predicting signatures of anisotropic resonance energy transfer in dye-functionalized nanoparticles†

Gabriel Gil,*[abc] Stefano Corni,[b] Alain Delgado,[de] Andrea Bertoni[b] and Guido Goldoni[ab]

Resonance energy transfer (RET) is an inherently anisotropic process. Even the simplest, well-known Förster theory, based on the transition dipole–dipole coupling, implicitly incorporates the anisotropic character of RET. In this theoretical work, we study possible signatures of the fundamental anisotropic character of RET in hybrid nanomaterials composed of a semiconductor nanoparticle (NP) decorated with molecular dyes. In particular, by means of a realistic kinetic model, we show that the analysis of the dye photoluminescence difference for orthogonal input polarizations reveals the anisotropic character of the dye–NP RET which arises from the intrinsic anisotropy of the NP lattice. In a prototypical core/shell wurtzite CdSe/ZnS NP functionalized with cyanine dyes (Cy3B), this difference is predicted to be as large as 75% and it is strongly dependent in amplitude and sign on the dye–NP distance. We account for all the possible RET processes within the system, together with competing decay pathways in the separate segments. In addition, we show that the anisotropic signature of RET is persistent up to a large number of dyes per NP.

## Introduction

Dye-functionalized nanoparticles (NPs) constitute a class of nanomaterials with great perspectives in a number of biological and medical applications as well as in optoelectronic devices,[1] due to the possibility to tailor specific electronic and optical properties by controlling the size and composition of the synthesized NPs.[2] The undergoing phenomena to be exploited are often related to light-induced excitations followed by intra-system processes, occurring between the dye and the NP, which compete with intrinsic decay pathways in the separated fragments.[3,4]

One of these processes is resonance energy transfer (RET) between the dyes and the NP.[3] An early description of this non-radiative mechanism was obtained by Förster in the context of dye–dye energy transfer,[5] whereby the rate of a transition between the excited state of the donor and the acceptor is estimated assuming an interaction between point-like transition dipole moments (TDMs). Although this approximation needs to be relaxed when the fragments have very different size and multipole contributions are relevant,[6] Förster theory nicely exposes the inherently anisotropic character of RET *via* dipole–dipole interactions. Indeed RET anisotropy is well established in the case of fixed dye–dye pairs (*e.g.*, terminally attached to DNA helices).[7–10] However, the anisotropic nature of dye–NP RET remains largely unnoticed either in theoretical or experimental accounts.[11–13] Note that, even in the case of spherical NPs, RET cannot be considered isotropic when the underlying lattice is anisotropic (*e.g.*, in wurtzite structures).[14] On the other hand, typical experimental setups average over possible orientations of the point dipoles, leading to an apparent isotropic RET rates.[15] This might not be the case when the large NP fragment of the dye–NP pair cannot reorient itself within the typical RET timescale and, *a fortiori*, when the dye is firmly anchored to the NP, so that the relative orientation is fixed.

In this article we propose that the anisotropic character of RET could be exposed in the photoluminescence (PL) from properly designed hybrid nanomaterials composed of a NP decorated with several molecular dyes. Our theoretical strategy enable us to find very different PL spectra when the system is excited with orthogonal linear polarizations of the laser, which can be traced directly to the anisotropic character of dye–NP RET and the underlying crystal structure of the NP. Such anisotropy could be employed for practical applications, such as converting input light polarization in different output PL signals. In real systems, RET processes may take place between the NP and the dyes as well as between the dyes themselves. Since such RET rates can be comparable to those of intrinsic excitation (*i.e.*, absorption) and de-excitation (*i.e.*, radiative and non-radiative) processes, we develop a kinetic approach to estimate the steady state of the system in terms of the excited state populations of the dyes and the NP, whereby the full PL spectrum is obtained. A combination of state-of-the-art calculations and literature experimental data allow us to find magnitude and direction of the transition electric field (TEF) generated by the NP and of the dye TDM, from which RET rates are determined. We use, in particular, a generalized Förster theory[6] where all multipole contributions in the TEF of the large NP are taken into account. We show that the anisotropy depends on the number of dyes anchored to the NP's surface, due to competing dye–NP and dye–dye RET. However, we estimate that the anisotropy is robust with respect to the dispersion in the number of dyes per NP in typical samples.

[a]S3, CNR-Istituto Nanoscienze, Via Campi 213/A, 41125 Modena, Italy. E-mail: gabrieljose.gilperez@nano.cnr.it; Tel: +39 059 205 5283

[b]Dipartimento di Scienze Fisiche, Informatiche e Matematiche, Università degli Studi di Modena e Reggio Emilia, Via Campi 213/A, 41125 Modena, Italy

[c]Instituto de Cibernética, Matemática y Física, Calle E No. 309, 10400 Havana, Cuba

[d]Advanced Research Complex, University of Ottawa, 25 Templeton Street, K1N6N5 Ottawa, ON, Canada

[e]Centro de Aplicaciones Tecnológicas y Desarrollo Nuclear, Calle 30 No. 502 e/5ta y 7ma Avenida, Playa, Havana, Cuba

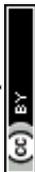

## System description and anisotropic RET

We focus on a core/shell CdSe/ZnS wurtzite (hexagonal lattice) semiconductor NP with spherical symmetry (core and total radii of 2.3 and 2.8 nm, respectively)[16] coated with an arrangement of six uniformly distributed cyanine dyes (Cy3B) and immersed in water. In Fig. 1, we show a schematics of the hybrid nanocomposite (NC) architecture.

A wurtzite CdSe/ZnS NP exhibits a two-fold degeneracy of the lowest excitonic state corresponding to orthogonal TDMs $\mathbf{d}_{NP}^{(\pm)}$ (with $\mathbf{d}_{NP}^{(+)} \perp \mathbf{d}_{NP}^{(-)}$) oriented along the 2D hexagonal lattice of the crystal.[17–19] These NP states cannot be directly excited by linearly polarized light if the polarization direction is orthogonal to both TDMs. The plane defined by $\mathbf{d}_{NP}^{(\pm)}$ is called "bright plane", whereas the direction normal to it is called "dark axis" (see Fig. 1). Ultimately, the latter is the cause of the anisotropy in dye–NP RET since there is no other source of directional symmetry breaking in our architecture. Cubic lattice (*e.g.*, zinc-blende)[20] NPs with three-fold degeneracy of TDMs should not present this kind of anisotropy.

We assume that the dyes are rigidly linked to the NP with a given orientation, and with the TDM corresponding to the first absorption peak, $\mathbf{d}_M$, oriented radially with respect to the NP (Fig. 1).[21] DNA strands, *e.g.*, are good candidates as rigid linkers[22] in dye–NP RET setups,[23] when the dye/NP fragments are terminally attached to them. Notably, DNA linkers allowed experimental measurements of dye–dye RET anisotropy.[7] The linker (*e.g.*, DNA itself) can be selected in a way that it does not affect the excitation spectra of the NP or the dye.[24] Since it does not participate in any of the possible RET processes it is neglected in our modelling. We have selected our NC such that the dye is the donor, and the NP is the acceptor for the excitation energy in a dye–NP pair.[25] Although this is opposite to the most investigated role assignment in energy transfer setups based on dye–NP nanohybrids,[3,26] it is critical to expose the RET anisotropy. Dye–dye RET may be affected by phenomena akin to PL blinking and bleaching;[27,28] for dye–NP RET these phenomena should be even more rare.[29] PL blinking (on/off) time windows in the order of a second have been documented for dyes[30,31] and NPs.[29,32,33] We expect our analysis to hold during 'on' time lapses, since the kinetics of the excited states populations will be dominated by absorption, emission and RET processes occurring in a completely different timescale (nanoseconds). In any case, blinking and bleaching was not hindering single-NC dye–NP RET experiments before.[34,35] Finally, we suppose that an ensemble of NCs is deposited on a sample surface in order to fix the dipole directions in space. Of course, on the surface NCs may have random position and orientations. Moreover, we assume that the deposited NCs have a sufficiently low surface density and thus do not interact with each other. Then, the orientation of the dark axis for each NC separately can be determined by analyzing the polarization orientation of its PL,[17] and then one can focus on a single NC with the proper orientation (see below).

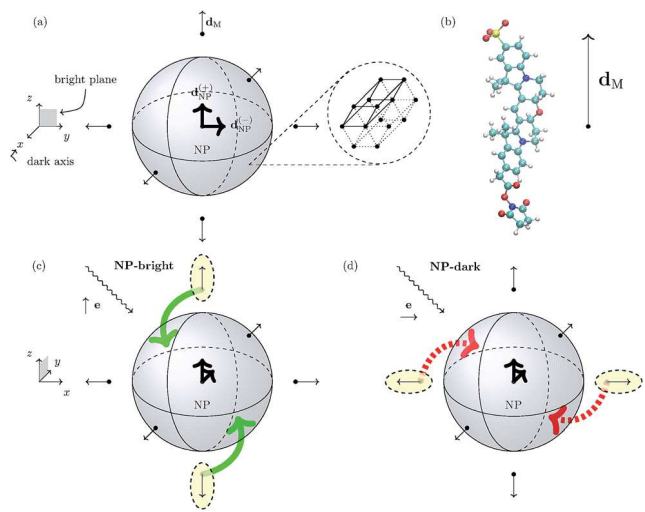

**Fig. 1** Panel (a) Schematic representation of the spherical NP with a zoom-in of its 3D hexagonal lattice unit cell. The peripheral arrows in the NC sketch indicate the orientation of the TDM of the dye molecules ($\mathbf{d}_M$), whereas the central arrows represent the degenerate TDMs of the ground state exciton in the NP ($\mathbf{d}_{NP}^{(\pm)}$). Panel (b) Schematic representation of the Cy3B dye molecule along side the orientation of the TDM of the excited state of interest in the internal reference frame.† Panels (c) and (d): Pictorial representation of dye–NP RET processes in two excitation conditions: (c) NP-bright, *i.e.*, $\mathbf{e} \| \mathbf{d}_{NP}^{(\pm)}$, where RET is allowed (indicated with green solid arrow), and (d) NP-dark, *i.e.*, $\mathbf{e} \perp \mathbf{d}_{NP}^{(\pm)}$, where RET is prohibited (red dashed arrow). The dyes that are excited in each case are indicated with yellow ellipses. Note the different axes orientation in (a) and in (c) and (d).

When a NC is irradiated with a linearly polarized laser, dye excitation occurs with a probability $\propto |\mathbf{e} \cdot \mathbf{d}_M|^2$, where $\mathbf{e}$ is the polarization vector of light, so that dyes with $\mathbf{d}_M \| \mathbf{e}$ are preferentially excited. Excited dyes have two main decay pathways: (i) intrinsic decay, either radiative (fluorescence) or non-radiative (dissipation to a thermal bath), and (ii) RET to close acceptors, either another dye or the NP. Although all these processes may have comparable rates, it is possible to tune the number

and position of dyes (as we shall show in the following), so that dye–dye RET is almost suppressed with respect to intrinsic mechanisms, while dye–NP RET is dominating, at least for those dyes which are correctly aligned with the NP dipoles.

In turn, the dye–NP RET rate, $k_{M \to NP}$, depends on the dye position. Due to the anisotropic dipole–dipole interaction, only dyes with a component of $d_M$ in the bright plane can transfer energy to a NP, while those with $d_M$ parallel to the dark axis cannot (Fig. 1). Therefore, if the NC is irradiated with $\mathbf{e} \| d_{NP}^{(+)}$ (or equivalently $\mathbf{e} \| d_{NP}^{(-)}$) energy can be transferred to the NP directly from the laser through absorption and indirectly through dye–NP RET. We call this polarization "NP-bright" configuration. If, on the contrary, the NC is irradiated with $\mathbf{e} \perp d_{NP}^{(\pm)}$, both mechanisms are inhibited.[36] We call this polarization "NP-dark" configuration. The two configurations are shown in Fig. 1(c) and (d). We shall now discuss how PL can be used to reveal RET anisotropy.

For a dye at a generic angular position (setting the NP center as the origin of coordinates), with $d_M$ components both in the bright plane and along the dark axis, dye–NP RET is viable even in the NP-dark configuration. However, since RET is a fast function of the dye–NP distance ($\propto$(distance)$^{-6}$ within Förster theory), we can tune the distance so that the average dye–NP RET rate (weighted with absorption rate), exceed the rate of intrinsic decay of the dye, $k_M$, for NP-bright configuration, while the opposite is true for NP-dark configuration. In this case, the most efficient decay path for the excited dyes is RET for NP-bright, whereas intrinsic de-excitation pathways is preferred for NP-dark.

The PL of the NC is the combination of radiative de-excitation of the NP and all dyes on its surface. After the laser excitation at the lowest absorption frequency of the dye, for NP-dark configuration, light is re-emitted mainly at the PL frequency of the dyes, because the NP does not absorb light and part of the excited dyes cannot transfer energy to the NP. For NP-bright configuration, instead, we expect enhanced PL from the NP due to both direct absorption and efficient RET from the excited dyes. In the latter case, PL of the dye would correspondingly be quenched. Therefore, PL difference between NP-dark and NP-bright configurations is a direct fingerprint of an anisotropic dye–NP RET. Note that in order to probe the predicted anisotropic character of the dye–NP RET, one should first establish the NP-bright and NP-dark orientations in a single NC, which stem from the analysis of the polarization orientation of PL of the NP.[17,18]

## Excitation, de-excitation and RET rates

To evaluate quantitatively the modulation of the PL spectra as a function of laser polarization, we setup a kinetic model for the excited state populations of dyes and the NP which includes explicitly the competition between (i) intrinsic intra-system decay mechanisms, (ii) direct RET (dye → NP), (iii) back-transfer (NP → dye), and (iv) diffusion of the excitation in the molecular shell through dye–dye RET. The level-alignment of the LUMO orbital of the dye and the bottom of the conduction band of the NP, in principle allows also for electron transfer mechanisms at short distances ($\sim$0.5 nm from the surface of the NP).[37] Evaluating the competition between energy- and charge-transfer is beyond the scope of this article. Hence, we estimate our results to be valid when the latter process is negligible, i.e., for dye–NP center-to-center distances larger than $\sim$4.5 nm (nearest atom-to-surface distance of $\sim$0.5 nm) in our particular case of radially oriented Cy3B.

### Absorption and intrinsic decay rates

The absolute absorption rate could be calculated from the intensity and frequency of the excitation laser. For our purposes, we take an estimated value of the dye absorption rate, i.e., $k_{abs}^M \approx 0.005$ ns$^{-1}$. We get the ratio between absorption rates from the experimental ratio of extinction coefficients of the NP and the dye at the absorption peak of the dye, i.e. $k_{abs}^{NP}/k_{abs}^M = 1.340$ (Table 1). Therefore, $k_{abs}^{NP} \approx 1.340 \times 0.005 \approx 0.007$ ns$^{-1}$.

The theoretical fluorescence rate of the dye is[17,38]

$$k_{fl}^M = \frac{4}{3}(\Delta E_M \alpha)^3 |d_M|^2, \tag{1}$$

where $\Delta E_M$ is the dye de-excitation energy, and $\alpha$ the fine structure constant. $k_{fl} = 1/\tau_{fl}$, where $\tau_{fl}$ is the characteristic (radiative) time of fluorescence, can be computed from the radiative quantum yield $Q_y = \tau/\tau_{fl}$, where $\tau$ is the lifetime of an excited state, whether decaying from radiative or non-radiative pathways. Hence, we can recover the dye TDM from $\Delta E_M$, the quantum yield and the lifetime of the dye excited state, namely, $Q_y^M \tau^M$ and $\tau_M$.

In Table 1 we show these figures in the case of Cy3B (dye)[39,40] and the NP.[16] The NP radiative lifetime $\tau_{fl}^{NP} \sim 5$ ns is an estimate from values in the literature for similar quantum dots,[37] leading to a total decay lifetime of $\tau_{NP} = Q_y^{NP} \tau_{fl}^{NP} \sim 1.6$ ns (Table 1). Thus, the total intrinsic decay rate of the dye and NP are $k_M = 1/\tau_M = 0.357$ ns$^{-1}$ and $k_{NP} = 1/\tau_{NP} = 0.625$ ns$^{-1}$, respectively. The de-excitation energy is $\Delta E_M = hc/\lambda_{fl}$.

### RET rates

In general, the RET rate is given by[5,41,42]

$$k_{DA} = 2\pi |V_{DA}|^2 J_{DA}, \tag{2}$$

where $V_{DA} \equiv \langle D^*, A | \hat{V} | D, A^* \rangle$, $\hat{V}$ is the donor–acceptor interaction Hamiltonian, and $|D\rangle$ and $|D^*\rangle$ ($|A\rangle$ and $|A^*\rangle$) the ground and the excited electronic states of the donor (acceptor) fragment. $J_{DA}$ is the spectral overlap between the normalized donor emission ($f_D(E)$) and the acceptor absorption ($a_A(E)$) spectra,[43,44]

$$J_{DA} = \int dE f_D(E) a_A(E). \tag{3}$$

Table 1 Cy3B dye and NP parameters used in the kinetic modeling, namely, wavelengths for absorption and fluorescence maxima ($\lambda_{abs}$ and $\lambda_{fl}$), quantum yields of radiative decay ($Q_y$), extinction coefficients ($\varepsilon$), lifetimes ($\tau$), and the computed TDM ($d$). $\varepsilon$ correspond to $\lambda_{abs}$. Experimental data for the dye is taken from ref. 39 and 40, while for the NP we use ref. 16

|  | Cy3B | NP |
|---|---|---|
| $\lambda_{abs}$ (nm) | 558 | — |
| $\lambda_{fl}$ (nm) | 572 | 606 |
| $Q_y$ | 67% | 32% |
| $\varepsilon$ (M$^{-1}$ cm$^{-1}$) | 150 000 | 200 943 |
| $\tau$ (ns) | 2.80 | $\sim$1.6 |
| $d$ (D) | 11.95 | — |

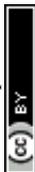

For the dye–dye RET, we can employ the Förster dipole–dipole theory since the dyes are sufficiently far apart, i.e.,

$$V_{M_1 M_2} = \frac{-\left(3(\mathbf{d}_{M_1} \cdot \mathbf{R}_{M_1 M_2})(\mathbf{d}_{M_2} \cdot \mathbf{R}_{M_1 M_2}) \big/ R_{M_1 M_2}^2 - \mathbf{d}_{M_1} \cdot \mathbf{d}_{M_2}\right)}{\varepsilon R_{M_1 M_2}^3}, \tag{4}$$

where $\mathbf{R}_{M_1 M_2}$ is the vector connecting the molecules $M_1$ and $M_2$, which are point-like in our model. $\varepsilon$ is the optical-frequency dielectric constant[45] of water accounting for screening effects due to the dynamic component of the solvent polarization. All the renormalization of the dipole–dipole interaction due to the mismatch between dielectric environments inside and outside the dye volume is contained in $\mathbf{d}_M$, and hence not explicitly present in the interaction Hamiltonian.

For a quantitative analysis of dye–dye RET rates we choose a specific configuration for the shell of dyes tethered to the NP. We place each dye at a vertex of a regular octahedron concentric with the spherical NP (Fig. 1). We also assume that the dyes (six of them in total) are attached to the NP coherently with its crystal symmetry. Therefore, two of them have $\mathbf{d}_M \| \mathbf{d}_{NP}^{(+)}$, other two $\mathbf{d}_M \| \mathbf{d}_{NP}^{(-)}$, and the remaining two $\mathbf{d}_M \perp \mathbf{d}_{NP}^{(\pm)}$ (Fig. 1). In the octahedron configuration, the distance between neighboring dyes is $R_{M_1 M_2} \equiv R_{MM} = \sqrt{2}R$, where $R$ is the dye–NP distance from the NP center to the center of charges of the dye. In experiments $R$ is usually determined by the length of the molecular chain linkers.

In the case of dye–NP RET, we account for the very different size of the two segments which is not taken into account in the Förster dipole–dipole coupling. We used a recently developed generalization that accounts for the interaction between all transition multipole moments of the NP and the transition dipole of the dye.[6] These interactions capture non-monotonic behaviour of the RET rate with $R$ that is overlooked by Förster approximation, and extends its validity to small distances. In this scheme, the electronic coupling is written as

$$V_{M,NP} = -\mathbf{d}_M \cdot \mathbf{E}_{NP}(\mathbf{R}), \tag{5}$$

where $\mathbf{E}_{NP}(\mathbf{R})$ is the TEF generated by the NP at the position $\mathbf{R}$ of the center of charge of the dye.[46]

The topmost valence band of wurtzite CdSe NPs has a strong heavy hole character, while lower bands may show strong heavy/light hole band mixing effects.[47] Here, we consider only the bright ground state heavy-hole-like exciton, since higher excitonic states are either outside the energy range allowed for the transfer (defined by the partner dye PL spectrum), or they are not optically-active due to angular momentum selection rules.[47]

Due to a degeneracy of the interband bulk dipole moment in hexagonal crystal systems (i.e., the case of CdSe/ZnS wurtzite materials) we have two channels for the energy transfer, i.e.,

$$k_{M \to NP(NP \to M)} = 2\pi\{|\mathbf{d}_M \cdot \mathbf{E}_{NP}^{(+)}(\mathbf{R})|^2 + |\mathbf{d}_M \cdot \mathbf{E}_{NP}^{(-)}(\mathbf{R})|^2\} J_{M \to NP(NP \to M)}, \tag{6}$$

where $\pm$ superscripts label the TEFs coming from the two-fold degenerate NP states having the interband bulk dipoles oriented in z and y axes, respectively. The excitation dark axis is thus placed in the x direction. $M \to NP$ ($NP \to M$) tags identify the RET from dye (NP) to NP (dye). Notice that, although in the back-transfer ($NP \to$ dye) case the electronic coupling part is identical to the direct-transfer (dye $\to$ NP) one, the spectral overlap in general differ, i.e., $J_{M \to NP} \neq J_{NP \to M}$, leading to different RET rates, i.e., $k_{M \to NP} \neq k_{NP \to M}$.

For the calculations, we have used experimental figures whenever available, e.g. the absorption and PL spectra,† the quantum yields, and the lifetimes.[16,39,40] The TDM of the dye, the dye–dye and dye–NP spectral overlaps were directly computed from these quantities. Nevertheless, due to a continuum of excitations of increasingly high absorbance for short wavelengths in the NP, the normalization of the absorption spectrum, entering in the dye–NP spectral overlap calculation, is not well defined. We thus resorted to fit the excitation spectrum of the NP with a linear combination of two Gaussians, representing the first two peaks in the theoretical absorption spectrum (see below). The linewidths of both peaks and the height of the first peak were considered as fitting parameters, whereas the height ratio of the first and the second peaks, taken from theoretical calculations, was imposed as a constraint.

The excited states and the TEF of the NP, i.e., $\mathbf{E}_{NP}^{(\pm)}$ of eqn (6), are computed by a configuration interaction approach for an electron–hole pair in a spherical NP.[6] The screening due to the dielectric mismatch between the semiconductor and the surroundings is fully taken into account.[48] We consider the semiconductor media as homogeneous with an effective dielectric constant equal to the volume-weighted average of the dynamic dielectric constants of CdSe core and ZnS shell materials.†

In Table 2 we show our results for each of the above mentioned RET processes. Analyzing the lifetime RET/intrinsic decay ratio, we see that all of these mechanisms are present in our system for the closest distance considered, i.e., $R = 4.23$ nm, since $\tau_{DA} \ll \tau_D$.

### *Ab initio* calculation of the dye absorption spectrum

Although experimental absorption spectrum and TDM of the dye are employed to compute dye–dye and dye–NP RET rates, we perform theoretical calculations to: (i) assess the dielectric screening effects of the NP in the dye excitation spectrum, and (ii) obtain the direction of the TDM in the internal reference frame of the dye. To that aim, we perform Time-Dependent Density Functional Theory (TDDFT) calculations for the dye. To account for the dielectric polarization effects of the solvent and the NP on the ground and excited states of the dye, we use the Polarizable Continuum Model (PCM).[45,49]

**Table 2** RET/intrinsic decay lifetimes ratio and RET rate for each of the possible RET process. The dye–NP distance selected was $R = 4.23$ nm

| D–A | $\tau_{DA}/\tau_D$ | $k_{DA}$ (ns$^{-1}$) |
|---|---|---|
| Dye–NP | 0.002 | 177.812 |
| NP–dye | 0.071 | 8.846 |
| Dye–dye | 0.286 | 1.249 |

We optimized the structure of Cy3B in vacuum through a DFT calculation using the B3LYP exchange–correlation (xc) functional and the 6-31G basis set. The excited state computations in vacuum and in water were carried out by means of linear response TDDFT with PCM, using the CAM-B3LYP xc functional and the 6-31G(d,p) basis set. Corrected linear response calculations[50] is used to include the state-specific solvent response in the excited state corresponding to the first peak in the absorption spectrum. All the calculations were performed using a local version of the General Atomic and Molecular Electronic Structure System (GAMESS-US)[51] interfaced with an external code that builds the PCM matrices in the case of non-homogeneous dielectric environment composed by the solvent and a spherical NP in the proximity of the dye.[52] The optimized geometry in vacuum and the TDMs in vacuum, and in water with or without a near NP, are given in the ESI.†

In Fig. 2 we show the dye absorption spectra obtained in water, either isolated or in the presence of the NP. The latter is plotted for $R = 4.2$ nm, which is the minimum dye–NP center-to-center distance, corresponding to the dye cavity in contact with the NP. We compare with the experimental results obtained in water solution.[39] Note that the theoretical first peak of absorption in water is blue shifted by nearly 100 nm ($\sim 0.5$ eV) with respect to the experimental spectrum. This result lies within the typical accuracy of a TDDFT vertical excitation energies (i.e., 0.2–0.5 eV),[53] which is mostly dependent on the xc functional. Although this points out the lack of spectroscopic precision of our CAM-B3LYP calculations, a benchmark of the available xc functional to describe the Cy3B excitation spectrum is not within our aims. In fact, we can still rely on the internal consistency of TDDFT calculations with dielectric surroundings to assess the effect of the NP medium in the spectrum. From Fig. 2 it is apparent that the latter effect is negligible, as the dye absorption spectra in water with or without a near NP are essentially coincident.

The TDM of the excited state which corresponds to the first peak in absorption is plotted in the dye internal reference frame in Fig. 1. We found no significant differences in the magnitude or even direction of the $\mathbf{d}_M$ computed in water with or without a near NP; the angle between the latter two vectors is about 1″.† The theoretical value $d_M = 11.64$ D (for the isolated dye in water) is very close to the experimental figure of 11.95 D (Table 1).

# PL signature of anisotropic RET

### Kinetic model and steady-state spectra

We are now in the position to set up a system of coupled master equations to study the kinetics of these processes in detail. Absorption, intrinsic decay pathways, as well as all relevant RET processes are considered in the dynamics of excited states population. For the dye we consider the ground and the first excited state, while for the NP we consider the ground and the two-fold degenerate excited states characterized by $\mathbf{d}_{NP}^{(\pm)}$. We recall that we are interested only in the condition for which $\mathbf{e}$ is oriented in the $xz$-plane; particularly in the NP-bright and NP-dark orientations. The RET rates $k_{M,NP}$, $k_{NP,M}$ and $k_{MM}$ depend upon $R$, and consequently also the excited states populations. The set of master equations are written in the ESI.†

From the steady-state populations of the dye and the NP in the NP-dark and NP-bright orientations, we can build up the respective NC PL spectra as

$$I_{b/d}(\lambda, R) = c_{NP}^{b/d}(R)I_{NP}(\lambda) + c_{M}^{b/d}(R)I_M(\lambda), \qquad (7)$$

where $I_{b/d}(\lambda, R)$ is the PL intensity of the NC, $I_M(\lambda)$ ($I_{NP}(\lambda)$) the normalized PL intensity of the dyes (NP) in the absence of the NP (dyes), and $\lambda$ the emission wavelength. The coefficients $c_M^{b/d}(R)$ and $c_{NP}^{b/d}(R)$ are proportional to the steady-state populations of the dye and the NP, respectively. b(d) superscript corresponds to the NP-bright (NP-dark) configuration. Hereafter we focus in particular on the PL intensity at the dye fluorescence wavelength, namely $I_{b/d}(R) \equiv I_{b/d}(\lambda_{fl}^M, R)$.

### PL enhancement/quenching due to orthogonal excitation polarizations

Let us define a figure of merit that quantify the relative difference between NP-bright and NP-dark PL intensities at the dye fluorescence wavelength, i.e.,

$$q(R) = \frac{I_d(R) - I_b(R)}{I_d(R)}. \qquad (8)$$

$q(R)$ represents the relative quenching of the PL intensity as a function of the distance $R$ in passing from the NP-dark to the NP-bright configuration. Therefore, $q = 0$ for $I_d(R) = I_b(R)$ (no relative quenching), while $q = 1$ for $I_b(R) = 0$ (total quenching). Notice that $q$ takes negative values when $I_b(R) > I_d(R)$ (relative enhancement instead of quenching). We remark that $q$ is not representative of the PL intensity change with respect to the independent fragments (dye and NP) PL. The calculated $q$ vs. $R$ is shown in Fig. 3, while in the inset we show the PL spectra for NP-bright and NP-dark orientations calculated at the dye–NP distance of maximum quenching. Remarkably, the dye PL quenching can reach ~75%, showing that indeed the relative difference in PL due to RET anisotropy is substantial, which is the main result of this work.

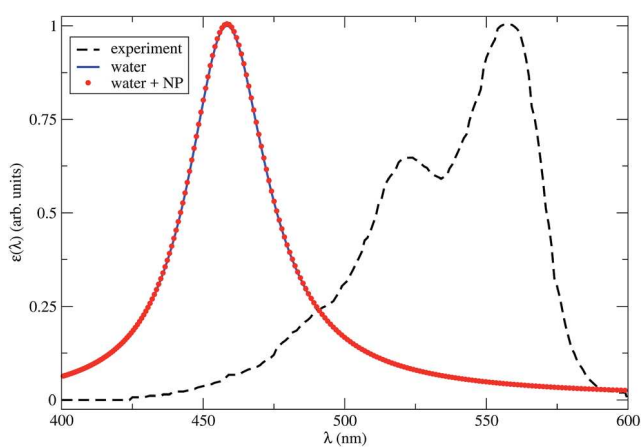

**Fig. 2** Theoretical Cy3B absorption spectra in water with a close NP (red dots), $R = 4.2$ nm, or isolated (blue solid line), calculated by TDDFT within the CAM-B3LYP approximation for the xc energy. The line broadening parameter is 0.1 eV. The experimental results in water solution (black dashed line) are also shown.[39]

We identify two distinctive regimes from Fig. 3. At very short dye–NP distances ($R \lesssim 4.15$ nm) $q < 0$, dye-related PL exhibit a relative enhancement for NP-bright configuration. This is unexpected on the basis of the previous reasoning. In this regime, for NP-dark configuration, there is a leak of dye excited state population into the NP via dye–dye and subsequent dye → NP RET processes (Fig. 4). For NP-bright configuration, instead, the NP can be directly excited via light-absorption, contributing to the population of the NP excited state. Increasing such population reduces the probability of dye → NP RET (blocking effect) and enlarges the probability of the opposite process (back-transfer effect). As a result of both effects,† the reduction of the dye PL is less severe in the NP-bright than in the NP-dark configuration, leading to the above mentioned relative enhancement. Indeed we have checked that if NP absorption is switched-off in the kinetic equations, such relative enhancement vanishes (see ESI† for a discussion).

For larger distances $q > 0$, dye-related PL shows a quenching in the NP-bright configuration. In this regime, in fact, dye–dye RET is no longer efficient, being dye–NP RET (available only for NP-bright) and intrinsic decay the only possible de-excitation pathways for the dyes (Fig. 4). Hence, for NP-dark configuration, dye-related PL is identical to the case of independent fragments. While, for NP-bright configuration, we observe the prescribed quenching of dye PL, when energy is transferred to the NP. At sufficiently large distances ($R \approx 20$ nm), the dye–NP RET rate is asymptotically vanishing, and the fragments become independent from each other. Thus, the PL spectrum is a superposition of that of the dyes and the NP. Furthermore, in this condition, the dye PL in NP-bright or NP-dark polarizations do not show any difference, as expected. Maximum PL quenching takes place at distances where dye → NP RET rate is much larger than intrinsic decay pathways. In the investigated NC, this happens at $R \approx 7.67$ nm (Fig. 3). At this center-to-center distance, the atom of the dye which is closer to the NP is located about 3.7 nm from its surface. Note that it is the dye–dye RET rate dependence on $R$ which controls the range of distances for which the largest quenching of the PL is taking place.

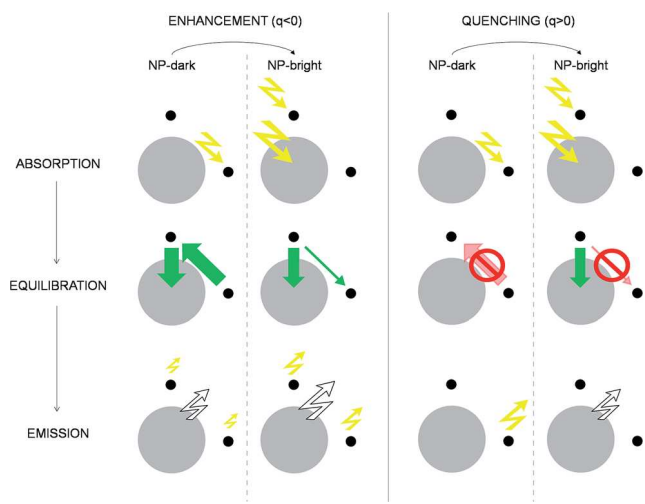

**Fig. 4** Excitation/de-excitation kinetics (absorption → equilibration → emission) for the enhancement/quenching regimes in the NP-bright/NP-dark configurations, as indicated. The NP is represented with a gray circle and dyes with smaller black circles. For the sake of simplicity only $xz$-section is drawn, and symmetrical molecules are removed. Green arrows indicate active RET processes. Inhibited RET processes are illustrated with transparent red arrows and a prohibition sign. Zigzag arrows stand for absorbed (incoming) and emitted (outgoing) light by the NC. In particular, yellow arrows refer to absorption (emission) at the dye absorption (fluorescence) wavelength, while white arrows mark the emission at the NP PL wavelength. The width of the arrows indicate qualitatively the likelihood of the process. See ESI† for quantitative information.

To emphasize that the PL enhancement/quenching is the result of a dynamical equilibrium between different mechanisms, we sketch in Fig. 4 the processes taking place in a possible PL experiment in the two regimes. The absorption process is intrinsically anisotropic, since the NP is only excited in the NP-bright orientation. In the equilibration stage, a balance between dye → NP, NP → dye and dye–dye RET mechanisms is established. In the relative enhancement regime, for NP-dark, there is an efficient pathway to deliver dye photoexcitations to the NP through intermediate dyes, while, for NP-bright the direct excitation of the NP reduces the net flux of excitation from the dye to the NP, thanks to both blocking[54] and back-transfer effects as discussed above. Thus, the overall effect is a larger excited state population of the dyes in the NP-bright with respect to the NP-dark configuration, and, ultimately, less PL intensity in the NP-dark configuration. In the quenching regime, due to inactive diffusion of the excitation in the dye shell, no transfer is allowed in the NP-dark, and the dye-related PL matches that of the independent fragments, while, for NP-bright, efficient dye → NP RET overcomes back-transfer to decrease the dye excited state population, ensuing finally, a quenched dye-related PL.

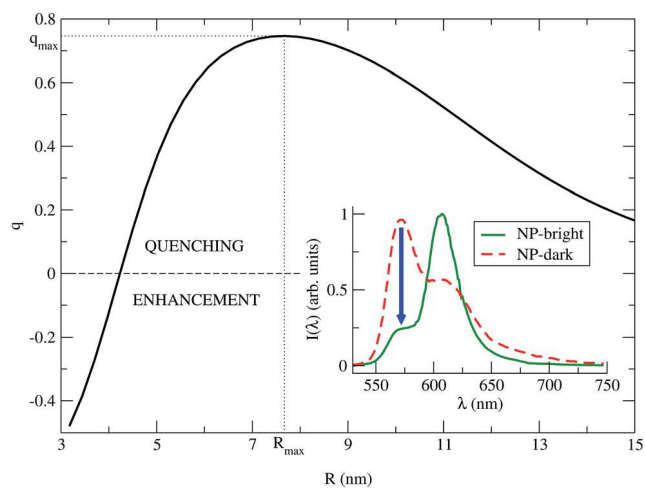

**Fig. 3** $q$ vs. $R$ is plotted for $\lambda = \lambda_{\mathrm{fl}}^{\mathrm{M}} = 572$ nm (eqn (8)). The maximum quenching $q_{\max} = 0.75$ is found for the distance $R_{\max} = 7.67$ nm. Inset: PL spectra of the NC (eqn (7)), for NP-bright (green solid) and NP-dark (red dashed) polarizations, with $R = R_{\max}$. The blue arrow indicate the maximum quenching in the spectra.

Although both the enhancement and the quenching regimes are clear signatures of anisotropic dye–NP RET, in the following, we focus on the quenching since the relative enhancement occur at very short distances where (i) the absolute difference of NP-bright and NP-dark dye PL intensities is very low† and thus unsuitable for experimental verification and (ii) electron-transfer, not considered in our kinetic model, may have an important effect on the PL. Moreover, our discussion of the mechanism leading to relative enhancement highlighted the decisive role of the NP absorption that is trivially anisotropic, while our goal is to expose the anisotropy of the RET process itself.

## PL quenching vs. number of dyes per NC

While we have explicitly solved above a set of equations for a few dyes tethered at selected positions, in practice it may be difficult to control the exact number of dyes that are attached to each NP. Indeed, during the synthesis only the average concentration can be tuned, and the number of dyes per NC follows a Poisson statistics.[37]

To evaluate qualitatively the effect of changing the number of dyes per NP, $N$, without the need to solve ad hoc kinetic equations, we identified the two main conditions that lead to $q > 0$ (i.e., PL quenching in the NP-bright polarization). These are: (i) the rate of the dye → NP RET in the NP-bright polarization should be larger than the intrinsic decay rate $k_M$ of the dye; (ii) the rates of the dye–dye RET and the dye → NP RET in the NP-dark polarization are both smaller than $k_M$. In fact, under these conditions, PL from the dye will be quenched (with respect to the case of independent fragments) through the dye → NP process in the NP-bright configuration while it will not be affected in the NP-dark configuration.

An estimate of such rates can be obtained by defining average dye–dye and dye–NP RET rates $\Phi_{MM}^{b/d}$ and $\Phi_{M \to NP}^{b/d}$. $\Phi_{MM}^{b/d}$ represents the rate of dye–dye RET averaged on all the dyes and taking into account the excitation probability of each dye; $\Phi_{M \to NP}^{b/d}$ is the analogous quantity for the dye–NP RET. Their expressions in terms of the RET and absorption rates are given in the ESI.† Notice that the average dye–dye RET rate $\Phi_{MM}^{b/d}$ is also dependent on the polarization (b or d) due to the way we are constructing the dye shell (i.e., here the symmetric octahedron tessellation; † less symmetric shapes as hexagonal prisms[14] can also be considered). We will use henceforth $\Phi_{MM} \equiv (\Phi_{MM}^b + \Phi_{MM}^d)/2$ in the analysis bellow.

Fig. 5 shows $\Phi_{MM}$, $\Phi_{M \to NP}^b$ and $\Phi_{M \to NP}^d$ as a function of $R$. $k_M$ is the reference line below which processes are not likely to occur. The different rates are plotted for the three representative cases $N = 6$, 18 and 38. We remark that, for $N > 6$, the dye shells considered here† imply that some dyes have a TDM that is neither orthogonal nor parallel to the NP dark axis. $N = 6$ has been treated quantitatively before, providing a benchmark for the present qualitative treatment. In this case, $\Phi_{M \to NP}^d \ll k_M$ for $R$ larger than the NP radius (2.8 nm), and therefore $\Phi_{M \to NP}^d$ is not shown in Fig. 5. Notably, there is a range of dye–NP distances, i.e., ~6–12 nm, for which the two conditions required to find a quenching in the NP-bright PL emission are met (represented by the inset in the figure). Such range compares reasonably well with the quantitative results of Fig. 3, where quenching is found for $R > 4$ nm and $q_{max}$ is around 7.7 nm. For $N = 18$ it is still possible to find a range of $R$ corresponding to quenching (red shadowed area in the inset), and the optimum distance is about ~11 nm. In the latter case, the quenching range is smaller than for $N = 6$ since $\Phi_{MM}$ is closer to $\Phi_{M \to NP}^b$, while $\Phi_{M \to NP}^d$ is irrelevant in both cases. In other words, what limits the PL quenching is the excitation diffusion inside the dye layer rather than the level of anisotropy of dye–NP RET. Finally, for $N = 38$, $\Phi_{MM} > \Phi_{M \to NP}^b$ and the PL quenching is lost due to such excitation diffusion. There is thus an upper limit on the number of dyes for NP that can be used to reveal RET anisotropy via the present PL (or, from another perspective, to convert a change of light polarization in a change of PL intensity and spectral shape).

To approximately quantify the maximum quenching for a given $N$, we can use a simplified steady-state model. Based on the discussion just presented, we assume that excitation diffusion in the dye shell and the NP-dark dye–NP RET are inactive processes since the maximum quenching is reached under these conditions. Moreover, we neglect the back-transfer from the NP, and we assume an excitation regime far from the saturation of the excited state population upon absorption or RET. Under these assumptions, $q_{max}$ reads

$$q_{max} = \frac{\Phi_{M \to NP}^b}{k_M + \Phi_{M \to NP}^b}, \qquad (9)$$

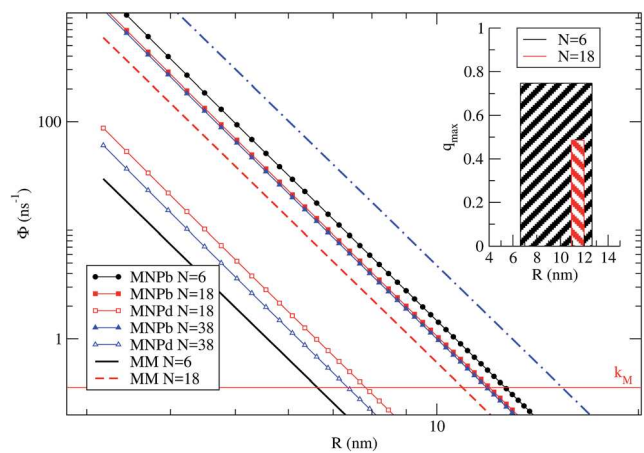

**Fig. 5** Average dye–dye (MM) and dye–NP (MNP) RET rates as a function of $R$, for $N = 6$, 18, and 38. The reference horizontal line mark the intrinsic decay rate of the dye, $k_M$. NP-bright and NP-dark configurations are marked with letter 'b' and 'd' in the legend. Inset: representation of the $R$ window for which it is possible to find a quenching in NP-bright PL emission (w.r.t. NP-dark) in the cases of $N = 6$ and 18. The bar width indicates the optimal $R$ range, while the height correspond to the quenching value (eqn (9)) at the middle point of the interval.†

*i.e.*, $q_{max}$ is directly related to the quantum yield for dye de-excitation *via* dye–NP RET in the NP-bright configuration. For $N = 18$, $q_{max}$ reduces to ∼65% the value corresponding to $N = 6$ (Fig. 5 inset). We also note in Fig. 5 that the range of distances for PL quenching should be narrower and centered at higher values when increasing $N$ (Fig. 5 inset).

Finally, we have verified that for a concentration of dyes such that $N = 6$, the average quenching would not be affected by a realistic (*i.e.*, Poissonian) dispersion on the number of dyes per NC (see ESI† for details).

## Concluding remarks

In summary, for properly designed dye-functionalized NPs, we predict substantial difference in dye PL for two orthogonal linear polarizations of the excitation laser. The relative difference of the dye-related PL intensities between the two polarizations can be as high as 75% for low number of dyes per NP. The maximum difference is found at a dye–NP distance large enough to rule out the role of dye–NP electron transfer. Moreover, we find that the change in dye PL with the incident light polarization is still observed up to a large number of attached dyes per NP, although the maximum relative difference in PL intensities is expected to move to larger distances, and to decrease in amplitude as the number of dyes increases. More importantly, we have shown that the difference in PL is a clear signature of the anisotropic nature of dye–NP energy-transfer, and it should be detectable in single-NC micro-PL spectroscopy.

Finally, we comment on the advantageous use of radially oriented TDMs of the dyes in our prototypical system. We have chosen the latter since it preserves spherical symmetry, while other orientations, such as tangential, are less symmetric, and would lead to further anisotropies in the dye–NP RET not related to the inherent anisotropy of the NP. The donor/acceptor role assignment to the dye/NP is also convenient due to the fact that, in such a way, PL from the donor do not include an anisotropic absorption bias.

## Acknowledgements

We acknowledged financial support from European Community's FP7 Marie Curie Initial Training Network project INDEX (Grant Agreement No. PITN-GA-2011-289968). S. C. acknowledges that this work has received funding from the European Research Council (ERC) under the European Union's Horizon 2020 research and innovation programme (Grant agreement No. 681285 – TAME-Plasmons). We thank J. I. Climente for useful discussions on the excitonic ground state of wurtzite CdSe NPs.

## Notes and references